# Planar pinning induced, lowering of vortex dimensionality and low field melting in a single crystal of Ba$_{0.6}$K$_{0.4}$Fe$_2$As$_2$

**Ankit Kumar[1], Sayantan Ghosh[1], Tsuyoshi Tamegai[2], S. S. Banerjee[1*]**

[1]Department of Physics, Indian Institute of Technology, Kanpur-208016, India

[2]Department of Applied Physics, The University of Tokyo, Hongo, Bunkyo-ku, Tokyo 113-8656, Japan

*email: satyajit@iitk.ac.in

**Abstract**

Theoretically, the vortex melting phenomenon occurs at both low and high magnetic fields at a fixed temperature. While the high field melting has been extensively investigated in high $T_c$ cuprates, the low field melting phenomena in the presence of disorder hasn't been well explored. Using bulk magnetization measurements and high-sensitivity differential magneto-optical imaging technique, we detect a low-field vortex melting phenomenon in a single crystal of Ba$_{0.6}$K$_{0.4}$Fe$_2$As$_2$. The low field melting is accompanied with a significant change in local magnetization ~ 3 G, which decreases with increasing applied field. The observed vortex melting phenomena is traced on a field temperature phase diagram and which lies very close to theoretically predicted Lindemann criteria based low field melting line. Our analysis shows a Lindemann number $c_L$ = 0.14 associated with the low field melting. Imaging of low-field vortex melting features shows the process nucleates via formation of extended finger like projections which spreads across the sample with increasing field or temperature, before entering into an interaction-dominated vortex solid phase regime. Magnetization scaling analysis shows that the dimensionality of melting vortex state is close to one. Angular dependence of bulk magnetization hysteresis loop in our sample shows the presence of extended defects. From our studies, we propose the sample contains a peculiar geometry of extended defects arranged in a plane in the sample, with these planes extending through the sample thickness. In the weak intervortex interaction limit, we argue that reduced vortex dimensionality due to pinning by these peculiar extended defect planes strongly enhances thermal fluctuations. It is these extended defects planes, which we propose are promoting low dimensional vortex melting in the pnictide system.

**Introduction**

A conventional, ordered atomic lattice with long range positional order exhibits a first order thermal melting into a liquid phase. Presence of quenched random disorder in the lattice leads to loss of positional order thereby suppressing the first order nature of this melting transition. The vortex state in type II superconductors is a convenient prototype for studying the behavior of phase transitions in a soft condensed matter system in the presence of thermal fluctuation and pinning effects. In the context of vortices, their pinning by defect and impurity sites in a superconductor, is technologically important as the pins immobilize vortices driven by electric currents, thereby reducing the vortex flow induced dissipation. Pinning leads to loss of long range order in a vortex solid [1]. Thermal fluctuations often counter pinning effects by thermally activating vortices out of the pinning centers thereby effectively weakening the pinning potential. In the soft vortex state, the competition between intervortex interactions trying to generate an ordered vortex configuration and pinning and thermal fluctuations trying to destroy long-range order in the vortex state, leads to variety of different static vortex matter phase [1,2,3,4,5,6,7,8,9,10,11,12,13,14]. In pinned vortex solids the positional correlations between vortices decay either algebraically or exponentially [1,11,13,14], and they are called Bragg or Vortex glass phases respectively. The elastic moduli of the vortex solid are magnetic field ($B$) dependent [1]. At low $B$ when the intervortex spacing $a_0 > \lambda$ (where, $\lambda$ is the superconducting penetration depth, $\phi_0$: magnetic flux quantum), the elastic moduli of the lattice are small due to weak intervortex interactions. Due to non-local effects, the elastic moduli also decreases at high $B$ where $a_0 \ll \lambda$. At intermediate magnetic field strengths where the elastic moduli take maxima, the vortex solid phase is stabilized. Softening of elastic moduli triggers a thermally-induced melting phenomenon in the vortex solid. Thermal fluctuations acting on a soft vortex solid melt it into a vortex liquid phase, wherein the positional correlations exist only between nearest neighbor vortices. Conventionally, a popular criterion to describe melting is the Lindemann criterion [1], $\sqrt{\langle u^2 \rangle} \sim c_L a_0$, where $\sqrt{\langle u^2 \rangle}$ is the r.m.s. deviation of the vortex line from its equilibrium position due to thermal fluctuations, and $c_L \sim 0.1 - 0.2$ is the empirical choice of the Lindemann number. Using such a Lindemann criterion and softening of the elastic moduli of the vortex solid, the boundary of stability of the vortex solid phase in a field ($B$) - temperature ($T$) phase diagram is derived. It was shown that in a pinning-free system, the vortex solid melts into a vortex liquid phase at both high and low $B$ [3,1,15]. Thus, a unique characteristic of the $B$ - $T$ vortex matter phase diagram is that, the phase boundary across which, the continuous symmetry of the vortex liquid phase is broken as it forms a vortex lattice, is encountered not only at high but also at low $B$. High $T_c$ superconductors (HTSC) with their enhanced $T_c$, large anisotropy, and their complex vortex structure comprising of string of interacting two dimensional vortices [1,14], became popular systems for studying the vortex solid to liquid melting phenomena [6,7,8,9,11,16,17] at high $B$. At high $B$, the vortex solid to liquid transformation was shown to be like an ice to water like transformation wherein the vortex liquid is relatively denser compared to the solid phase (vortex density, $\rho = B/\phi_0$) [18,19]. Conventionally it is believed that the presence of quenched random disorder leads to loss of positional correlations in the vortex lattice thereby obliterating all evidences of a melting phenomena [20,21,22]. While overall melting features can be understood via the Lindemann criterion, the vortex dimensionality also plays an important role in governing this phenomenon. It has been proposed that melting from a one-dimensional (1D) stack of two dimensional (2D) vortices into a 2D vortex liquid phase occurs as enhanced thermal fluctuations either overcomes the coupling between the 2D vortices in the stack [20] or by disrupting the net intervortex interactions within a stack [22]. While numerous studies exist on investigating vortex solid melting phenomenon at high vortex densities (high $B$), comparatively fewer studies exist on dilute vortex melting. As per conventional understanding, pinning effects should dominate at low fields, hence the dilute vortex solid at low fields is likely to be a strongly pinned and configurationally disordered. Hence one may ask, in such disordered vortex solids, is there a chance of vortex melting at low fields? Experimentally it is not quite well established if a pinned vortex solid melts at low $B$ in realistic samples with pinning. Some recent studies in clean HTSC at low fields have suggested the presence of high vortex mobility regions and changes in local vortex density at low $B$ [23,24]. However, apart from HTSC which have a complex vortex structure, there exists no evidence of melting at low $B$ in any other materials. It may be mentioned that in the dilute regime melting signatures are masked by the presence of strong magnetization irreversibility induced by strong pinning effects. In fact, at low fields, theoretically, the presence of a pinned glassy vortex phase has been proposed in the past, viz., 'the reentrant glass' [12].

In recent times pnictides class of superconductors have been extensively investigated. These materials possess moderately high $T_c$'s, small superconducting coherence length ($\xi$), and moderate anisotropy, which makes the vortex state in these materials potentially susceptible to thermal fluctuations. Theories have predicted vortex solid melting phenomenon in different pnictide materials, like, LaFeAsO$_{1-x}$F$_x$, Ba(Fe$_{1-x}$Co$_x$)$_2$As$_2$ and Nd(O$_{1-x}$F$_x$)FeAs [25, 26]. A source of complication in pnictide materials towards experimentally observing signatures of thermal melting of the vortex solid is that they usually possess very strong pinning. It is known that pinning induced irreversibility masks signatures related to vortex melting phenomenon, such as changes in equilibrium magnetization associated with a change in vortex density as the vortex solid transforms into a liquid [27]. Most imaging studies on the vortex state in pnictide superconductors report a disordered vortex solid which persist up to high $B$ [28,29], confirming the presence of strong pinning in this class of superconductor. Studies suggest the presence of microscopic chemical inhomogeneities in the system to act as point pinning centers in the material [30]. However, a recent study in K doped 122 systems, viz., Ba$_{0.6}$K$_{0.4}$Fe$_2$As$_2$ have shown the presence of an ordered vortex solid present at high field [31], where presumably the intervortex interactions have managed to overcome the vortex pinning strength in this material. It has also been reported that $Ba_{0.5}K_{0.5}Fe_2As_2$ shows thermodynamic signatures of vortex melting at high fields [32]. In this paper, we explore signatures of a vortex liquid phase at low *B* in a single crystal of $Ba_{0.6}K_{0.4}Fe_2As_2$. We find that at low fields, while effective pinning is large, the pinning strength is distributed between weak and strong pins. At low *B,* we observe a change in equilibrium local magnetization ~ 3 G, which is associated with a melting transition. Using the high sensitivity differential magneto-optical imaging technique to spatially map up the location of changes in local *B,* we show low-field melting begins as linear finger like fronts projecting into the sample from different locations on the sample edge. Eventually, they spread all across the sample before merging into each other as *B* or *T* is increased. We show that due to nucleation of vortex liquid at the sample edges, the edge screening currents redistribute in a way that, partially it flows along the sample edge and part along the interface between vortex solid and liquid phases creating regions with suppressed vortex density inside the sample. The vortex melting transition is located in a *B-T* phase diagram and which lies in close proximity to the theoretically proposed low-field melting line based on a Lindemann melting criterion also plotted in the phase diagram. In the phase diagram, we also identify different low and high field vortex phases. Our explorations on the dimensionality of the melting vortices using scaling analysis of the bulk magnetization indicate that the dimensionality of the vortices is close to one. Our angular dependent magnetization study shows the presence of elongated defects extending along the sample thickness, which we believe is responsible for lowering the vortex line dimensionality to near one dimension. Based on our results, we propose $Ba_{0.6}K_{0.4}Fe_2As_2$ possesses a configuration of strong pinning linear defects arranged in a plane which extends into the sample thickness. We argue that the unusual configuration of an extended defect pinning plane in the crystal reduces the dimensionality of vortices, which in turn makes the vortices strongly susceptible to thermal fluctuations thereby precipitating the low-field melting phenomenon in this system.

**Experimental**

We report results on a single crystal of $Ba_{0.6}K_{0.4}Fe_2As_2$ with dimensions $1.7 \times 1.2 \times 0.025$ mm$^3$ and $T_c = 38$ K chosen from a batch grown using a self-flux method in $Al_2O_3$ crucibles [33]. In our experiments, the applied magnetic field ($B_a$) was maintained ‖ to the crystallographic c-axis. Bulk magnetization measurements were performed in a commercial Cryogenic SQUID magnetometer. For imaging the distribution of local magnetic field component (which is proportional to the local vortex density) ‖ to the crystallographic c-axis ($B_z$) across the sample, we use conventional magneto-optical imaging (MOI). We also use high sensitivity differential magneto-optical (DMO) imaging technique to measure changes in local vortex density ($\delta B_z$). Details of both MOI and DMO techniques are discussed elsewhere [21,34,35,36]. Briefly, in MOI we image the Faraday rotated light intensity distribution ($I(x,y)$) of linearly polarized light reflected from the sample, where $I(x,y) \propto B_z(x, y)$, note $B_z$ direction is ‖ $B_a$ and the co-ordinates $(x, y)$ are along the sample surface. In a DMO technique for the same $B_a$, we obtain an image which is the average of repeated differential image captured at $B_a$ = (MO images captured at $B_a+\delta B_a$) – (MO image at $B_a$) with $\delta B_a$ = 1 G. The intensity distribution in the differential image $\delta(x, y)$ in the DMO images was calibrated to map the changes in the local magnetic field $\delta B_z(x, y)$ produced at different locations inside the sample in response to external field modulation of $\delta B_a$ = 1 G.

**Results and discussion**

Figure 1(a) shows bulk magnetization hysteresis loop measurement at 35 K. The hysteresis loop width $\Delta M$ is related to the pinning strength experienced by the vortices inside the superconductor. Figure 1(a) shows that $\Delta M$ undergoes a modulation due to the presence of a second magnetization peak (SMP) anomaly in this sample. The SMP anomaly in magnetization have been seen in HTSC's [37,38,39,40,41,42,43,44,45] as well as in different low $T_c$ superconductors, for example in 2H-NbSe$_2$ [46,47] and also in Ba$_{0.6}$K$_{0.4}$Fe$_2$As$_2$ [48,49] apart from other iron-based superconductors [50,51,52,53]. Small angle neutron scattering investigations of the SMP anomaly in Ba$_{0.64}$K$_{0.36}$Fe$_2$As$_2$ shows this anomaly is associated with an order to disorder transition in the vortex state precipitated by proliferation of topological defects in the vortex state at high $B_a$ [49]. We would like to mention that all our investigations are performed in a $B_a$ regime which are far below the SMP regime which typically begins from 0.4 T (where there is a minima in $J_c^{bulk}$) and extends up to 1.2 T (where the $J_c^{bulk}$ peaks).

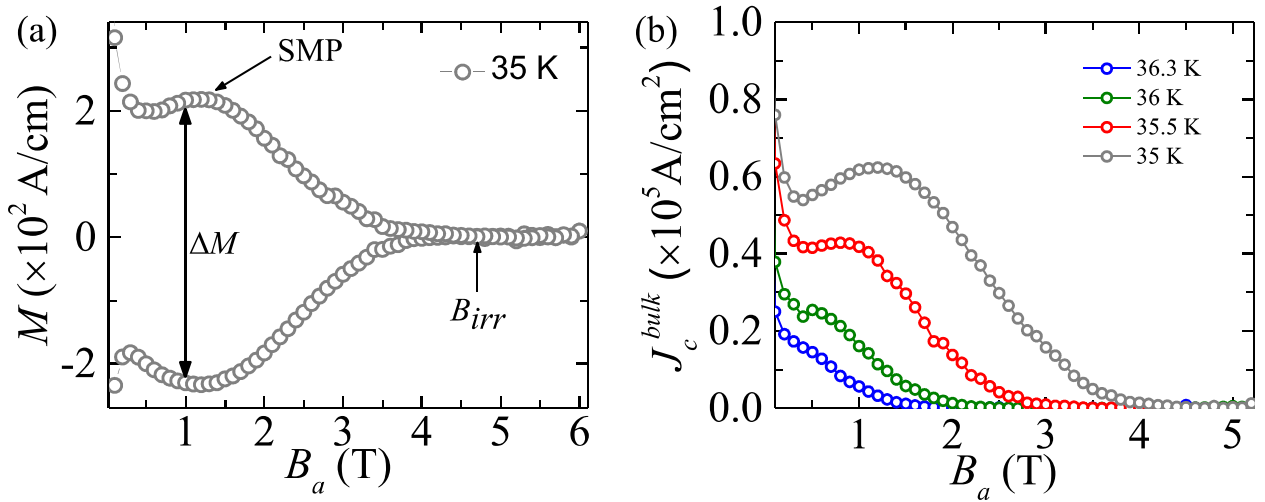

FIG. 1. **(a)** DC magnetization ($M$-$B_a$ curve) loop measured in a single crystal of Ba$_{0.6}$K$_{0.4}$Fe$_2$As$_2$ with $T_c$ = 38 K for applied magnetic field $B_a \parallel$ crystallographic $c$-axis at 35 K. The second magnetization peak anomaly (SMP) is seen to develop at $B_a \sim 0.4$ T and extends up to 1.2 T. Here we have chosen the symbol size comparable to the size of the error bars. **(b)** Critical current density $J_c^{bulk}$ vs applied field measured at different temperatures are plotted.

We estimate $J_c^{bulk}$ from $\Delta M$ (see Fig. 1(a)) using [54,55], $J_c^{bulk} = 20\Delta M/[a(1 - a/3b)]$, and $a$ and $b$ ($b > a$) are the crystal dimensions perpendicular to $B_a$. The $J_c^{bulk}$ ($B_a$) behavior overall has a monotonically decreasing trend with increasing $B_a$ as shown in Fig. 1(b) and also shows a broad bump associated with SMP. Note that the $J_c^{bulk}$'s value at low $B_a$ are high in the range of ~ $10^4$ A/cm² even at $T$'s > $0.9T_c$, where one expects large thermal fluctuations to suppress pinning effects significantly. This suggests the presence of strong pinning in the sample.

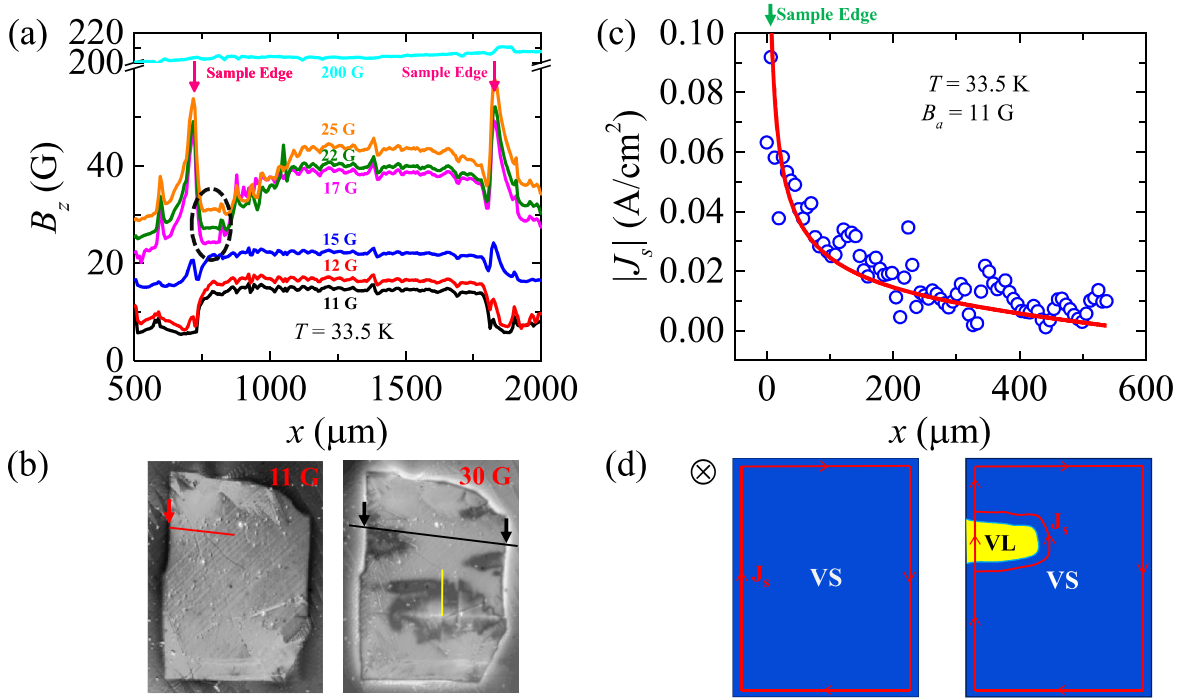

FIG. 2. **(a)** Figure shows the $B_z$ profile along the black line drawn in figure b (conventional MO image at 30 G) at 33.5 K and at different applied magnetic field from 11 G to 200 G. Note that, far away from the sample edges (beyond the scale shown in the figure) local field $B_z$ reduces to applied field $B_a$. Positions of the sample edges are indicated in the plot. Dashed black circular region in the plot shows the depletion of vortices density (or lowering of $B_z$). **(b)** Figure shows the MO images taken at 33.5 K and at an applied field of 11 G and 30 G respectively. **(c)** Screening current inside the sample obtained by inverting the conventional MO image acquired at an applied field of 11 G and temperature 33.5 K is plotted. Red curve shows the fitting through the data using equation $J_s(x) = B_a(w - 2x)/(d\sqrt{x(w-x)})$ (where $J_s$ is shielding current, $w$ is the sample width, $d$ is the sample thickness and $B_a$ is the applied magnetic field). Current distribution only in the half sample (see red line drawn in Fig. 2(b) in 11 G image) has been shown here with maximum current at the edge as shown in the figure. Here we have chosen the symbol size comparable to the size of the error bars. **(d)** The left schematic shows the screening current $J_s$ (red line) flowing at the edges of the sample in the presence of an applied magnetic field (⊗ mentions the direction of the magnetic field is going into the plane for both schematics). Right side schematic shows the formation of a vortex liquid puddle nucleated at the sample edge. The schematic on the right shows the shielding current redistributes by partially flowing along the sample edges and partially circulating along the boundary between the VL puddle and the VS phase. Note the schematic is only a representation relating to the formation of solid-liquid interface and the distribution of currents in this region (the schematic is not to scale). All the measurements have been done in the field cooled state with the magnetic field applied ∥ crystal c axis.

Figure 2(a) shows $B_z(x)$ measured across the line drawn in Fig. 2(b) (black line in 30 G image) at different $B_a$. The $B_z$ near the sample edges is seen to be enhanced due to strong shielding currents circulating on the sample edges and the $B_z(x)$ near the sample center exhibits a non-Bean like, dome-shaped profile. At high $B_a$ (see $B_z(x)$ for $B_a$ = 200 G in Fig. 2(a)), the uniform $B_z$ distribution across the sample is associated with the uniform vortex density in a vortex solid phase. Two conventional MO images of the sample at 33.5 K and 11 G and 30 G in the field-cooled (FC) state are shown in Fig. 2(b). Note that all measurements reported here are for the sample in a field-cooled state. Returning to the dome shape of the $B_z$ profile at lower fields, it is known that geometrical barriers induced presence of strong shielding currents near the sample edges [56] drive vortices away from the edges (where local $B_z$ is low) and they collect near the sample center. Consequently, the $B_z$ and hence the vortex density near the edges gets suppressed while the $B_z$ in the sample interior develops a dome like feature [56,57,58] (see Fig. 2(a)). In the $B_z(x)$ profile in Fig. 2(a), we see that between the sample center and near the left sample edge the $B_z(x)$ exhibits a significant dip. In fact, the region with suppressed $B_z$ expands in the direction of sample center with increasing $B_a$ (see within the black dashed circle in Fig. 2(a)). Note also that there is an asymmetry of this dip feature between the left and right sample edges. This dip feature shows that as the shielding currents get stronger with increasing $B_a$, resulting in significant number of vortices being pushed away from the sample edges. In Fig. 2(b) the MO image at 33.5 K and at an applied field of 30 G, shows the regions with suppressed $B_z(x)$ have a darker MO contrast compared to the surroundings. By numerically inverting [59] the measured $B_z(x,y)$ distribution at $B_a$ = 11 G and at $T$ = 33.5 K (Fig. 2(c)) we estimate the shielding current distribution $|J_s(x)|$ near the left edge of the sample (along the red line shown in Fig. 2(b) in 11 G image). For clarity, the current distribution only near one edge of the sample is shown in Fig. 2(c). The $|J_s(x)|$ in Fig. 2(c) fits the expression [56,60] for shielding currents, $J_s(x) = B_a\,(w - 2x)/(d\sqrt{x(w - x)})$, (where $w$ is the sample width, $d$ is the sample thickness and $B_a$ is the applied magnetic field). Using the fit, the shielding currents in the sample near the edges are $J_s \sim 10^{-1}$ A/cm² << $J_c^{bulk} \sim 10^4$ A/cm² (determined from bulk $M(B_a)$ measurements in Fig. 1(a), (b)). This apparent inconsistency is reconciled by considering the distribution of pinning strength in the sample is not uniform, and the sample has strong and weak pinning centers with the strong pinning centers dominating the bulk magnetization hysteretic response measurement. We believe these strong pinning centers which dominate the bulk $M(B_a)$ response are strong enough to survive up to high $T$ near $T_c$. On the other hand, the weak pinning centers have a low $J_c$ such that $J_s \sim 10^{-1}$ A/cm$^2$ is enough to depin vortices and push them away from these regions in the sample. The asymmetry in the suppressed $B_z$ feature between the left and right sample edges is related to non-uniformity of the pinning distribution across the sample. One possible cause for the observation of bright magneto-optical contrast over the finger like projections at 30 G (T = 33.5 K) in the MO image of Fig. 2(b), is related to enhanced density of trapped vortices pinned in strongly pinned regions of the sample. If this was a valid possibility then, we should have observed this brightening feature down to low fields like 11 G at 33.5 K for the FC state. However, in Fig. 2(b) we do not see the brightening features down to 11 G. Furthermore, if the brightening appearing in the MO image at 30 G in Fig. 2(b) was related to a strong pinning region, then such strong pinning regions should shield out modulations of external magnetic field. Hence the bright regions in Fig. 2(b) at 30 G should exhibit negligible changes in local field ($\delta B_z$) in response to a modulation of the external field (for example see Fig. 3 of Ref. [21]). To explore this feature, we perform differential magneto-optical imaging technique discussed below.

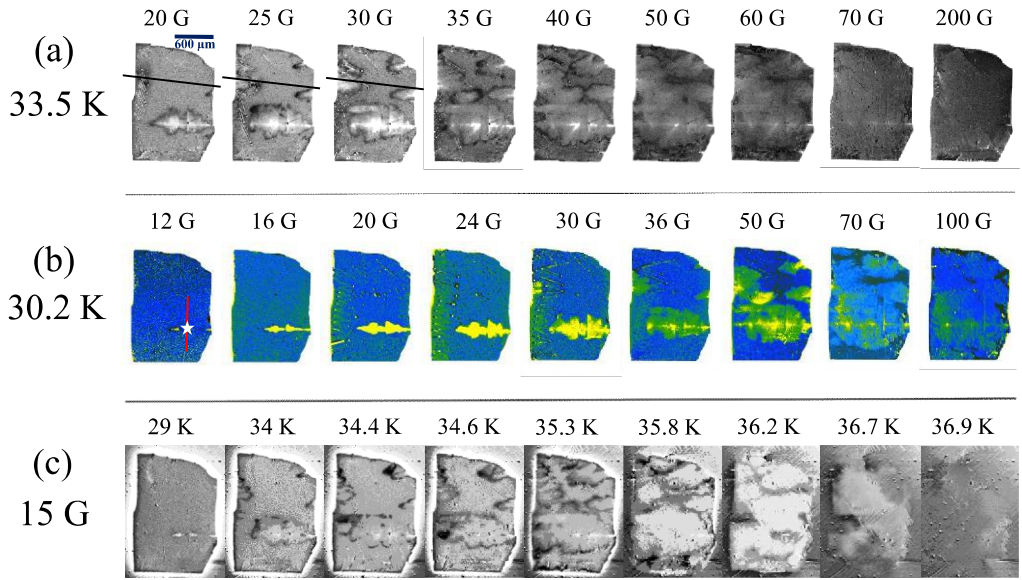

FIG. 3. **(a-b)** Show the differential MO (DMO) images taken at 33.5 K and 30.2 K respectively at different $B_a$. Figure b has been colored for better representation of the propagation of vortex melting across the sample. In some of the images in figure b, e.g., 20, 24, and 30 G images show a zig-zag pattern at the left edge of the sample. This zig-zag pattern is due to well- known Bloch walls seen on the magneto-optical film [61] which is placed on top of the sample for magneto-optical imaging. **(c)** Isofield differential MO images taken at a constant applied field of 15 G at varying

temperature. In figure (a)-(c) gray (blue in Fig. (b)) region represents vortex density region $\delta B_z = \delta B_a = 1$ G and bright (yellow in Fig. (b)) region represents enhanced vortex density region with $\delta B_z > 1$ G (vortex liquid; VL). Note that contrast in all images has been enhanced compared to the raw images for better visibility of the features. However, all quantitative analyses involving $B_z$ have been done using raw images. All the measurements have been done in the field cooled state.

Figures 3(a) and (b) show DMO images obtained as a function of varying $B_a$ at fixed $T = 33.5$ K and 30.2 K, respectively, while Fig. 3(c) shows DMO images at fixed $B_a = 15$ G captured at different $T$. The gray shade in the images (Fig. 3(a)) changing from a whitish to blackish shade, represents variations in $\delta B_z$. Initially, for low fields at 10 G (not shown in Fig. 3(a) panel, but this feature is seen in Fig. 3(b) at 12 G), the whole sample has an almost uniform gray (blue in Fig. 3(b)) intensity. The gray regions (blue in Fig. 3(b)) have $\delta B_z = \delta B_a = 1$ G, wherein the density of vortices follows the changes in the external magnetic field. As $B_a$ is increased, bright (yellow in Fig. 3(b)) finger-like regions begin invading the gray (blue in Fig. 3(b)) regions of the sample from different locations on the sample edges (for e.g. see 25 G and 36 G images in Figs. 3(a) and (b), respectively). Notice the bright regions in the DMO images occur in the same sample location where the brightening was seen in the 30 G conventional MO image in Fig. 2(b). These bright regions (yellow in Fig. 3(b)) are not symmetric patches but possess a directionality in their shape. Over the bright regions (yellow in Fig. 3(b)) in DMO images, the $\delta B_z$ is larger than 1 G, and hence in these regions the local vortex density has changed more in comparison to other neighboring gray regions. In the vicinity of the enhanced local field, i.e., $\delta B_z$ it may be noted that the $B_z(x)$ is non - Bean like with a dome shaped profile (see Fig. 2(a)). Prior to the brightening observed over regions of the sample in the differential images, the $B_z(x)$ profile shows an almost uniform field distribution over these regions, suggesting the feature is not associated with flux penetration. With increasing $B_a$, we see more such bright (yellow in Fig. 3(b)) linear fronts invading from the sample edges and expand after which they begin merging into each other at higher $B_a$, for e.g., see 40 and 50 G images in Figs. 3(a) and (b), respectively. As $B_a$ increases and the bright fronts spread across the sample, they also become less bright. Note that at high $B_a$ of 200 G (Fig. 3(a)) the $\delta B_z$ over the sample again becomes uniform, namely $\delta B_z \sim 1$ G $= \delta B_a$. At these relatively high fields recall that Fig. 2(a) shows the $B_z$ distribution across the sample is uniform, viz., the vortex state has a uniform vortex density. The absence of any significant gradients in $B_z(x)$ at 200 G suggests this state is an ordered vortex solid phase (see video 1 in the supplementary material). In the isofield DMO images in Fig. 3(c) we observe similar features of bright regions with enhanced $\delta B_z$ developing and propagating across the sample with increasing $T$ at constant $B_a$. Note from Fig. 3(c) that the sample disappears uniformly at 36.9 K, suggesting the $T_c$ is uniform across the sample and which in turn indicates the homogenous quality of the sample. We would like to mention that the bright regions appear at the same location in the sample at unique field and temperature values which are independent of whether the measurement performed is an isothermal or isofield run. This suggests the thermodynamic nature of the change seen in these FC state measurements. We would like to mention that we are able to observe these features in the DMO measurements only up to $0.4T_c$. For $T < 0.4T_c$, the signature for the above transformation in the vortex state is masked by the irreversibility in the sample (as pinning strength enhances with decreasing $T$). Also note that at any high $T$ and $B_a$ above 200 G, we do not observe any change in local $B_z$ appearing (we have measured up to 600 G). As mentioned earlier, strong pinning regions should exhibit dark contrast in DMO images due to their ability to screen external field modulation. Note that the regions where bright finger like projections appeared in the sample at 30 G in the conventional MO image in Fig. 2(b) is also the location where the DMO contrast becomes bright (see Fig. 3). The brightening of magneto-optical contrast in the DMO image of Fig. 3 relates to enhanced $\delta B_z$ in response to the external field modulation of 1 G. Thus the observed brightening in DMO signal is not correlated with enhanced shielding response. We conclude that the brightening in Fig. 2(b) at 30 G at the elevated temperature of 33.5 K is not related to strong pinning in those regions.

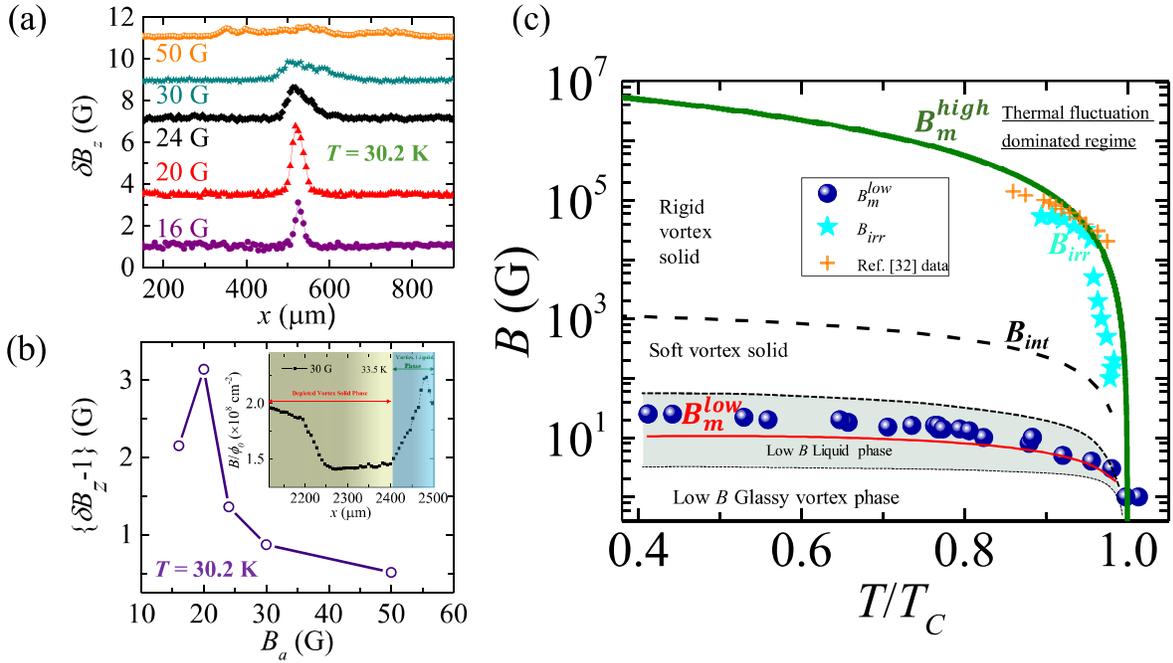

FIG. 4. (a) Local-field variation ($\delta B_z$) across the red line drawn in Fig. 3(b) is plotted (for 30.2 K). The $\delta B_z(x)$ plots are vertically shifted for the sake of clarity. Figure clearly shows with the increase in $B_a$ there is an enhancement in the local field present on the sample at a position marked as * as shown in the first figure of Fig. 3(b). (b) Shows the field dependence of the peak value of $\delta B_z$ above 1 G viz., the $\{\delta B_z - 1\}$ at the location marked with * on the sample in the first figure of Fig. 3(b). Inset shows the variation in the number of vortices/cm$^2$ (across yellow line in Fig. 2(b)) in the conventional MO image taken at 33.5 K and at applied field of 30 G. (c) $B(T)$ phase diagram (blue color data) is plotted for the sample region marked as * in the first figure of Fig. 3(b). The red color solid line is a fit to the low

field melting line ($B_m$), equation 1 (see text below). A shaded region, has been shown between two dashed line to distinguish between a liquid-like phase from a low field vortex glass and soft vortex solid phase at higher fields. Black dashed color line is the boundary of inter-vortex interaction $B_{int}(T)$ of the vortex matter. The olive color line is high field melting line. Cyan color data is the irreversibility data $B_{irr}(T)$ obtained from *M-B* curve. + symbol represent the high field melting data of $Ba_{0.5}K_{0.5}Fe_2As_2$ sample, from Ref. 32 Different phases have been identified in the different field regime of the phase diagram. In all the plots, we have chosen the symbol size comparable to the size of the error bars.

Figure 4(a) shows $\delta B_z(x)$ measured along the line drawn in Fig. 3(b) at different $B_a$ for the 30.2 K data, with the plots artificially offset for the sake of clarity. The appearance of the yellow regions at 16 G (see Fig. 3(b)) coincides with a change in $\delta B_z$ of 2 G, while $\delta B_z$ value is ~ 1 G (= $\delta B_a$) away from the yellow locations as discussed earlier in the context of Fig. 3. In Fig. 4(b), we plot the maximum value of ($\delta B_z$ -1) G versus $B_a$ at 30.2 K. Both Figs. 4(a) and 4(b) show that with increasing $B_a$, the magnitude of $\delta B_z$ over the yellow region changes by about 3 G at $B_a$ = 20 G before decreasing at higher $B_a$. Figure 4(a) shows the width of these regions also get broader with increasing $B_a$, which corresponds to the yellow regions spreading (see Fig. 3(b)). Above $B_a$ = 50 G the increase in $\delta B_z$ above 1 G reduces to 0.5 G and weakens further, which corresponds to a gradual weakening of expanding yellow regions as the vortex matter becomes denser at larger $B_a$. It appears that over the yellow regions, the dilute vortex state exhibits a larger change in local vortex density corresponding to a phase change in the vortex matter. At 30 G, we determine the behavior of $B_z(x)$ at 33.5 K in Fig. 2(b) over the vertical yellow line. The vertical yellow line in Fig. 2(b) extends from the uniform gray regions in the sample up to the bright regions where this same bright region in Figs. 3 and 4 exhibited an enhancement of $\delta B_z$ ~ 2 G (see Fig. 4(b)). Using the $B_z(x)$ determined above, in Fig. 4(b) inset we show that the brightening (yellow colored region) in Fig. 3(a) (in Fig. 3(b)) is associated with a change in local vortices density (= $B_z/\phi_0$). With increasing $B_a$ as the vortex density increases, the change in local field becomes smaller as shown in Fig. 4(b). The brightening associated with vortex density increase discussed above, we believe, is related to vortex melting phase transition wherein a dilute vortex state (VS) melts into a vortex liquid (VL) in a pnictide sample. Subsequently, we argue that the nature of the phase below the liquid at low fields as a disordered low density glassy vortex solid.

We argue here that further indirect evidence of formation of a vortex liquid phase is via the consideration that there is a redistribution of shielding currents in the sample due to the formation of VL phase which has relatively higher dissipation to flow of current (as vortex pinning is nominally zero here) compared to the VS phase which is a lower dissipation phase (with much higher pinning). Due to this, as the VL phase starts forming at the edges (Fig. 3), the shielding current divides and flowing partially along the sample edge and another flows along the VS-VL interface (see schematic in Fig. 2(d)). Note from Fig. 2(a) that the large $B_z$ near the sample edges shows that the shielding currents do not completely leave the sample edge when a VL puddle is nucleated at the edge. The shielding currents which circulate along the VS-VL interface in the sample interior are responsible for driving vortices away from the interface causing the observed depletion of vortex density at these locations (see dip feature in Fig. 2(a) and dark region in Fig. 2(b) in 30 G image). Around the edges of the VL puddle, we see the dark contours of the region with depleted vortex density in Fig. 2(b). We believe in these regions the sample has very weak pinning as vortices are driven away from the interface in these regions with shielding currents ~ $10^{-1}$ A cm$^{-2}$, (discussed earlier). As $B_a$ increases and the region with VL phase expands deeper into the sample, the region with lower $B_z$ (dark contrast around the bright region) moves deeper into the sample. Here we would like to mention that the black line drawn across the sample in Fig. 2(b) is at such a position that the suppressed $B_z$ (dark) region expands along the line for a few different $B_a$ values, with the bright region not crossing the line. Due to this in Fig. 2(a) we observed the suppressed $B_z$ region expand with increasing $B_a$. Note that other than the interface separating a VL-VS, it is unlikely to observe the above feature of current flowing along the interface causing vortex depletion. For example, consider an interface between two phases where in both phases the vortices are pinned albeit with different pinning strengths. In this case, as both phases are pinned, hence there is minimal difference in the resistivity of the two phases, hence currents would not have any reason to channel preferentially only along the interface.

In Fig. 4(c) we determine the melting phase boundary $B_m^{low}(T)$ correspond to the onset of brightening at the location marked by * in the first figure of Fig. 3(b). We use the criteria of a maxima in {$\delta B_z$ -1} (see Fig. 4(b)) to identify the location of $B_m^{low}(T)$ boundary at a given location in the sample. Around the $B_m^{low}(T)$, in Fig. 4(c) we draw two lines indicating the bounds of the region around the maximum in the {$\delta B_z$ -1}(B) curve in Fig. 4(b) where {$\delta B_z$ -1}>0 . We shade the region bounded between these two lines around $B_m^{low}(T)$, as the regime over which a phase transformation occurs. Later on we compare the $B_m^{low}(T)$ data with the theoretical predicted low field melting line equation. In the vortex melting phase diagram of Fig. 4(c) we also show the location of the reversible response of the superconductors (viz., irreversibility line, $B_{irr}(T)$), determined from the loss of hysteresis in bulk $M(B_a)$ measurements (see Fig. 1(a)). In a vortex matter phase diagram the location of the boundary across which thermal fluctuation dominate leading to vortex melting of the solid, is governed by the value of the Ginzburg number ($G_i$) [1]. For our K doped 122 sample we estimate $G_i = 1/2 \left[ k_B T_c \gamma / 4\pi B_c^2(0) \xi_{ab}^3(0) \right]^2 \sim 10^{-3}$ , using thermodynamic critical field $B_c(0) \left( \sim \frac{1}{\kappa} B_{c2}(0) \right) \sim 1.55$ T (reported [62] upper critical field $B_{c2} \sim 155$ T , $\kappa$

~ 100), ab-axis superconducting coherence length $\xi(0)$ ~ 1.2 nm [63] and the estimated anisotropy of our sample is $\gamma$ ~ 1.22 (see below). The $G_i$ value for our K-doped single crystal is between that of ~ $10^{-6}$ to $10^{-5}$ in low $T_c$ superconductors and ~ $10^{-2}$ to $10^{-1}$ in HTSC's. The relatively large $G_i$ suggests the vortices in this material are susceptible to thermal fluctuations effects. The olive-coloured line in Fig. 4(c) is the line obeying the formula for the high field melting of a vortex lattice for this pnictide system [1], given as $B_m^{high}(T) = (5.6 c_L^4/G_i) B_{c2}(0) \left(1 - \frac{T}{Tc(0)}\right)^2$ where $G_i = 10^{-3}$ and a standard Lindemann number $c_L$ of 0.2. We see the $B_m^{high}(T)$ line coincides with $B_{irr}(T)$, while the observed VS melting data points, $B_m^{low}(T)$ are well below $B_m^{high}(T)$ line. The melting phenomenon at high fields has already been studied in a K doped 122 Pnictide system very similar to ours, viz., in $Ba_{0.5}K_{0.5}Fe_2As_2$ single crystals, which has a $T_c$ slightly different from ours [32]. In the phase diagram of Fig. 4(c) the data from Ref. [32] is seen to lie on the theoretically predicted $B_m^{high}(T)$ line (shown as olive curve in Fig. 4(c)). In our phase diagram $B_m^{high}(T)$ line identified as the high field melting boundary across which thermal fluctuations completely overcome bulk pinning effects in the sample.

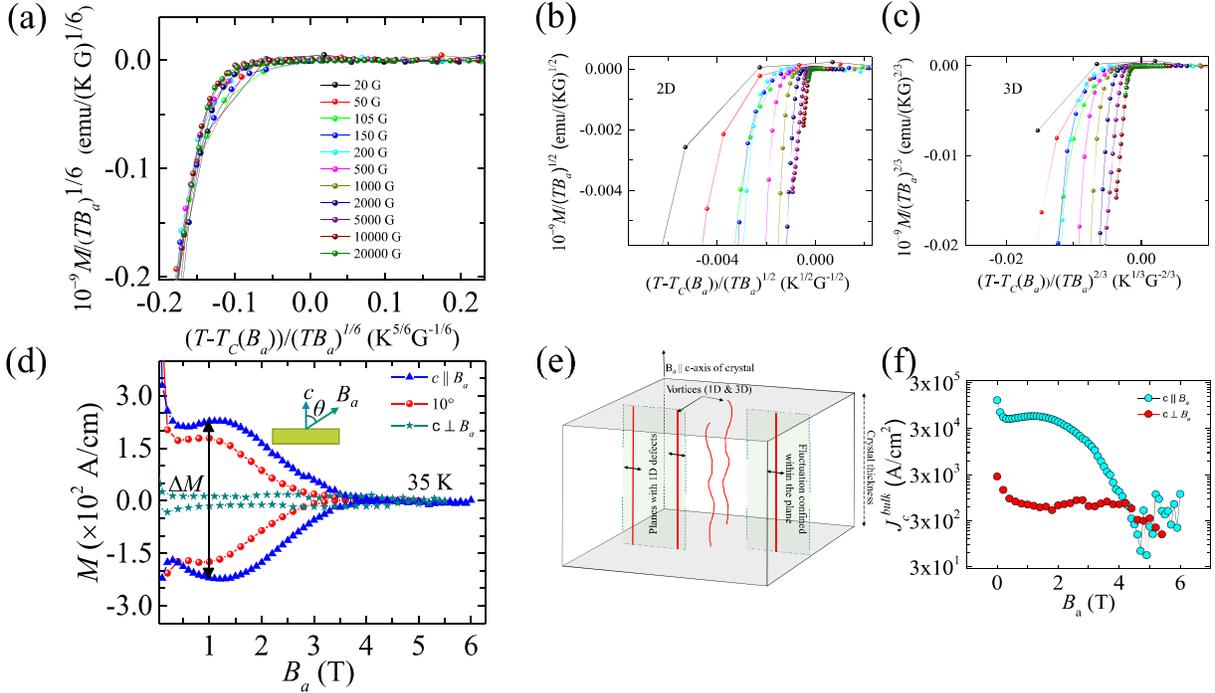

FIG. 5. (a) $M/(TB_a)^{1/6}$ versus $(T - T_C(B_a))/(TB_a)^{1/6}$ is plotted, obtained from $M$ vs. $T$ curves with different applied $B$ as shown in the figure (see text for details). (b-c) show scaling behavior is not followed for two and three dimensions. (d) Angular dependence study of the $M(B_a)$ hysteresis loop, which shows a decrease in the width of the loop with increasing angle with respect to c-axis as shown in the schematic. (e) Schematic shows planes of 1D defects inside the sample oriented along c-axis. (f) Critical current density $J_c^{bulk}$ vs applied field for orientation $B_a \parallel c$ and $B_a \perp c$ at 35 K. In all the plots from (a)-(f), we have chosen the symbol size comparable to the size of the error bars.

We next use scaling analysis used for HTSC [64,65,66,67,68,69], to investigate the dimensionality of the vortex state. In this analysis isofield $M(T)$ curves measured for our sample at low and high $B_a$ with the appropriate choice of dimensionality $D$, are replotted as $\frac{M}{(TB_a)^{(D-1)/D}}$ versus $\frac{(T-T_c(B_a))}{(TB_a)^{(D-1)/D}}$ [68,69] and all isofield $M(T)$ curves collapse onto a single curve. The scaling analysis we have performed is in a temperature regime close to Tc, viz., T/Tc(0) < 1%, where there is no irreversibility. Figure 5(a) shows $M(T)$ data from low to high fields are scaled by choosing $D = 1.2 \pm 0.1$. Figures 5(b) and (c) show the absence of scaling with D = 2 and 3. We believe that lowering the dimensionality of vortices in our sample towards one dimension is due to the presence of very strong pinning linear defects extending along the sample thickness. The slightly higher than 1D dimension seen in the above scaling analysis could be due to meandering of the vortex lines above 1D due to the contribution of pinning by weaker point defects, bringing the average dimensionality of above one. We believe that the fact the vortices have a dimensionality close to one due to the presence of extending pins in the sample. To search for evidence of extended defects in the sample, we perform angular dependent magnetization measurements. In Fig. 5(d) we see that the width of the hysteresis loop is maximum when $B_a \parallel c$ – axis of the single crystal and as we change the angle ($\theta$) between $B_a$ and crystallographic $c$ axis there is a significant decrease in the width of the loop. The loop width is minimum for $B_a \perp c$ ($\theta = 0°$). By incorporating demagnetization corrections, from the width of the irreversible magnetization loop measured for different $\theta$, we determine the $J_c^{bulk}$ vs $B_a$ for two different orientations, viz., $B_a \parallel$ to crystal c-axis (maximum demagnetization correction) and $B_a \perp$ to crystal c-axis (minimum demagnetization correction). For $\theta = 90°$ ($B_a \parallel$ to crystal c-axis) we use $J_c^{bulk} = 20\Delta M/[a(1 - a/3b)]$ for $b > a$, while for $B_a \perp$ to crystal c-axis with field gradient setup along sample thickness, $J_c^{bulk} = 20\Delta M/c$, where c is sample thickness. In Fig. 5(f) we see that at low fields, despite including demagnetization corrections the $J_c^{bulk}(\theta = 90°, B_a \parallel c) \sim 100 J_c^{bulk}(\theta = 0°, B_a \perp c)$. The significant change in $J_c^{bulk}$ as the orientation of the sample w.r.t to $B_a$ is changed suggests the presence of extended defects in the sample. The vortices align themselves with the defects extending along the sample c-axis (thickness) resulting in pinning in this orientation being maximum. The $M(B_a)$ in Fig. 5(d) shows that the high $J_c^{bulk}$ (large $\Delta M$) seen in Fig. 1 was due to extended strong pinning in the sample. The hysteresis loop width is much smaller in a direction perpendicular to these defects. We have shown earlier a preferential and directed nature of vortices penetrating into these defect planes (see supplementary information Fig. 1). This leads to higher vortex density and consequently a larger local field over these defect planes compared to surrounding regions. Due to this, the low field melting field value is reached first over these planar defects. Furthermore, the lowered dimensionality of the vortices in the planes makes them highly susceptible to thermal fluctuations effects leading to melting to begin as a linear puddle like feature in these defect planes. It may be worthwhile mentioning here that imaging of high field melting patterns in samples without such linear defects in BSCCO shows melting begins as a circular shaped liquid puddles and not as such linear shaped puddles [36]. Thus, the linear shape of the nucleated melted puddles in our samples we believe is a result of the presence of linear defect planes.

From the $M(B_a)$ loops measured in different orientations we plot the behavior of $B_{c2}(\theta)$ at 35 K (see supplementary information Fig. 2). The fit to $B_{c2}(\theta)$ with the known GL expression $B_{c2}(\theta) = B_{c2}(||c,T)(sin^2\theta + \gamma^{-2}cos^2\theta)^{-1\backslash 2}$ [70] gives an estimate of sample anisotropy $\gamma = 1.22 \pm 0.11$. It may be mentioned here that the anisotropy value, depends on doping levels in the sample and also on the temperature at which it is determined. Depending on the criteria used to determine $H_{c2}$ from either resistivity or magnetization measurements, there can be some spread in the reported anisotropy values. In similar samples like ours, the anisotropy at lower temperatures is about 2.6 [63].

At low fields (~ few tens Gauss), the intervortex spacing $a_0 \propto (\phi_0/B_a)^{\frac{1}{2}} \gg \lambda$ (~ 200 nm). In this regime, the vortices are weakly interacting as intervortex interactions in this regime go as $exp(-r/\lambda)$, where $r$ is the spacing between vortices. We believe that well below the $B_m^{low}(T)$ data in Fig. 4(c), at the low vortex density due to weak rigidity of the vortex state, there is a disordered glassy solid phase in our Pnictide superconductor with pinning. We call this as the low field glassy phase in Fig. 4(c) where this state can sustain a finite critical current at very low fields. In fact, we have stated earlier that below 0.4 $T_c$ the irreversible response of the vortex state at low fields is so large that we do not observe features of the change in $B_z$ associated with vortex melting. From this it appears that the low field glassy vortex phase can melt into a vortex liquid only at $T > 0.4\ T_c$. Below 0.4 $T_c$ the glassy vortex phase dominates the low field portion of the phase diagram and cuts $B_m^{low}(T)$ line where the thermal fluctuations are insufficient to melt the low field dilute vortex state. The Mermin-Wagner theorem [71] states that at finite temperature for a system with short range interactions in a dimensions lower than 2, it is highly susceptible to thermal fluctuations which prevent the spontaneous breaking of continuous symmetries in the system, viz., prevents ordering in such a system. Thus at finite $T$, in the low field regime around $B_m^{low}(T)$, where the vortices are very weakly interacting (as $a_0 > \lambda$) and in a dimension lower than two (as shown by the scaling analysis) viz., for the vortices confined in the plane, these vortices are strongly susceptible to thermal fluctuation effects. We would like to clarify that planar defects by themselves do not lead to increase in fluctuations. Rather the planar defects lead to the lowering of the vortex dimension thereby making the vortices susceptible to the thermal fluctuations in the low vortex density regime.

The lower dimensionality of the vortices helps in the triggering the melting phenomena to begin from the linear defect planes. As the vortex melting begins in these regions the elastic moduli of the vortex state in the surrounding regions decreases, thereby causing the melting to spread across the sample. Thus, we believe the weakly interacting vortices delocalized within the planes containing the linear defects (see schematic in Fig. 5(e)) are driven by strong thermal fluctuations, precipitating vortex melting at the low fields within the planes. This sort of melting produces the linear finger like projections, which occur over the location of the defect planes (Fig. 3(b) at 12 G). As $B_a$ is increased topological defects in the vortex state nucleating from this molten VL region begin to proliferate into the surrounding VS phase region thereby compromising the stability of VS region and subsequently melting it. Thus in this Pnictide system, we believe the defect plane induced reduction of vortex dimensionality triggers a low-field vortex melting near $B_m^{low}(T)$. In Fig. 4(c) we compare the $B_m^{low}(T)$ melting data points (blue solid circles) with the Lindemann criteria based low-field melting equation shown below [1],

$$B_m^{low}(T) \approx \frac{\phi_0}{\lambda^2}\frac{1}{4}[\ln(\frac{4\pi c_L^2}{(3\pi)^{1/4}}\frac{\epsilon_0 \lambda}{T})]^{-2} \quad , \tag{1}$$

where, $\lambda$ is penetration depth and $\epsilon_0 = (\phi_0/4\pi\lambda)^2$ is the vortex line energy [15]. In Fig. 4(c) the red line is a plot of eqn. (1) using $c_L$ = 0.14 [1]. For the plot we have used the penetration depth $\lambda(T)$ determined by fitting the lower critical field $B_{c1}(T)$ behavior we have measured for our sample using $\lambda(T) = \lambda_0 /(1 - (T/T_c)^2)^{1/2}$, with $\lambda_0 \approx 200$ nm (see supplementary information Fig. 3). In Fig. 4(c) we see that the theoretically predicted low field melting line and the experimentally determined $B_m^{low}(T)$ data closely follow each other with the theoretical curve being slightly lower in field compared to the experimental data. The vortex liquid phase lies below the upper boundary of the liquid phase and a vortex solid exists above it, and the theoretically predicted low field melting equation (1) seems to located in the center of the shaded liquid regime in the phase diagram of Fig. 4(c). At very low fields, however, there is a glassy vortex state as shown in our phase diagram. It may be remembered that the theoretically proposed low field melting line (equation (1)), is for an ideal pinning free system. We believe some of these differences from the theoretical low field melting boundary [1] and the experimental $B_m^{low}(T)$ data in Fig. 4(c) could be related to lowering of the vortex dimension induced by the presence of extended defects. It is surprising to find that although the dimensionality of the vortex state is reduced the VS to VL transformation is still reasonably well described by a Lindemann like form of eqn. (1). At very low fields in our sample, we have already suggested the presence of a low field glassy vortex phase. Around the low field melting $B_m^{low}(T)$ line which obeys the theoretically predicted low field melting line one has a liquid phase. At high fields with enhanced interactions between vortices viz., at $a_0 \sim (\phi_0/B)^{\frac{1}{2}} = \lambda(T) = \lambda(0)/(1 - (T/T_c)^2)^{1/2}$, the elastic moduli of the lattice get enhanced and the vortex solid forms across the sample. Using the above criterion of $a_0 = \lambda(T)$ we get the interaction boundary, $B_{int}(T) = \phi_0(1 - \left(\frac{T}{T_c}\right)^2)/\lambda(0)^2$, plotted in Fig. 4(c). Above the $B_{int}(T)$ boundary, the inter-vortex interaction dominates the behavior of the vortex matter. We believe that as the interaction dominated regime in the vortex matter phase diagram is approached the observed gradual diminishing of the brightness of the melting patterns in Fig. 3 is related to the VL transforming into a VS phase. In Fig. 4(c) above the shaded region around $B_m^{low}(T)$, there are no jumps in local $B_Z$, i.e., we observe uniform $B_Z$ across the sample (Fig. 2). Due to strong intervortex interactions setting in the vortex state only above $B_{int}(T)$, hence the vortex phase above the shaded liquid phase region and below $B_{int}(T)$ we call as a soft vortex solid phase with uniform gradients. Above $B_{int}(T)$ line in Fig. 4(c) we identify the rigid vortex solid which finally melts across the high field melting line, $B_m^{high}(T)$ and enters a fluctuation dominated regime.

It maybe noted that it is only above $B_{int}(T)$ theoretical boundary, where intervortex interactions dominate, we believe the denser quasi well order rigid vortex solid phase is expected to form. It is worthwhile noting that at high fields of above 200 G in Fig. 3(a), the linear finger like projections are no longer visible in the DMO images. We speculate that in the high-field intervortex interaction dominated regime viz., as $B_a \rightarrow B_{int}$, the effective pinning strength of the linear extended defects weakens. Note that the SMP anomaly in Fig. 1 is observed close to the $B_{int}(T)$ boundary. Here we estimate the difference in entropy ($\delta S$) between a glassy vortex state and vortex liquid phase. For the purpose of this calculation, in the low field regime where the intervortex interaction is weak, i.e., $B_m < B_{int}$ and $a_0 > \lambda$, we consider that the shear elastic moduli ($c_{66}$) of the vortex lattice goes to zero as the lattice melts. Then $\Delta U \approx c_{66} <u^2>$, is the difference in energy of thermally fluctuating vortices as the VS transforms into a VL phase [1,18] as $c_{66} \rightarrow 0$ in the VL phase and $\langle u^2 \rangle = c_L^2 a_0^2$ for melting. Around the melting line, the difference in entropy as one goes from the VS to VL phase is approximately, $\delta S \approx \frac{\Delta U}{T_m} \frac{\delta B_z}{B_z}$, where the temperature of the glassy vortex phase and of the vortex liquid phase near the melting line is approximated with $T_m$ and $\frac{\delta B_z}{B_z} \sim 2 \times 10^{-1}$ is the typical change in the vortex number density near the $B_m^{low}(T)$. As Pnictides possess $Fe_2As_2$ layered structure, it is convenient to estimate the entropy difference associated with vortex line segments of length $t$, where $t$ is the Fe$_2$As$_2$ layer spacing ($\sim$ 6.6 Å in our K-doped crystal). Using the above, the estimated δS between the dilute VS and VL phases and using $T_m$ = 22 K, is :

$$\delta S \sim \frac{c_{66}<u^2>t}{T_m}\frac{\delta B_z}{B_z} = \frac{c_{66}c_L^2 a_0^2 t}{T_m}\frac{\delta B_z}{B_z} = 0.0008 k_B,$$

where $c_{66} \approx \frac{\phi_0 B}{(8\pi\lambda(T))^2}$ [1,18], $c_L$ = 0.14, and $a_0 \sim (\phi_0/B_m)^{1/2}$. The difference in entropy between the low field glassy VS and VL phase is small (of the order of $0.001 k_B$). We believe we are able to detect this transformation at low fields due to the presence of the planar defects which lower the dimensionality of the vortices, thereby making them highly susceptible to thermal fluctuations effects. One may consider that with low field melting across $B_m^{low}(T)$, the effect of the pinning by planar pins should disappear. However, this consideration isn't always valid, especially in the presence of extended pins. Theoretical and experimental studies on melting phenomena in HTSC in the presence of low density of columnar defects show that the effects of pinning persist even into the high field liquid phase i.e., beyond the high field melting line, $B_m^{high}(T)$ [72]. Therefore, we believe that for these unusual defect planes the effect of thermal fluctuations on vortices is strongly enhanced along the planar defect planes. However, in a direction perpendicular to the defect planes the vortices remain confined, thereby retaining their one dimensional character deep in the fluctuation dominated regime. Consequently, we observe a one dimensional scaling valid deep in the thermal fluctuation dominated regime.

In summary, with this work, we have shown for the first time the existence of a low field vortex liquid phase transition in Ba$_{0.6}$K$_{0.4}$Fe$_2$As$_2$. We believe that it is the presence of peculiar configuration of extended defect planes in the sample reduces the vortex dimensionality, precipitating the vortex melting phenomenon. While pinning is known to destroy long-range correlations and suppresses evidence of a melting transition, there appears to be a unique planar pin configurations present in this pnictide system which helps to enhance effects of thermal fluctuations and drives a VS to VL melting transition rather than suppress it. We hope that our present work will stimulate further theoretical as well as experimental work in this direction to better understand the nature of pinning in these pnictides and the nature of the melting of the low dimensional vortices. Identification of such unusual defects in these pnictide crystals and their effect on pinning we believe has important ramifications, viz., to search for ways to control their configuration which in turn would help controlled dissipation in superconductor and thereby aid in high $J_c$ applications.

**Acknowledgement:** Satyajit S. Banerjee would like to acknowledge the funding support from IITK and DST-TSDP, Govt. of India.

**References**

1. G. Blatter, M. V. Fiegel'man, V. B. Geshkenbein, A. I. Larkin, and V. M. Vinokur, Rev. Mod. Phys. **66,** 1125 (1994).

2. A. A. Abrikosov, Zh. Eksp. Teor. Fiz. **32**, 1442-1452 (1957); [Sov. Phys. JETP **5**, 1174-1182 (1957)].

3. D. R. Nelson, Phys. Rev. Lett. **60**, 1973 1988; D. R. Nelson and H. S. Seung, Phys. Rev. B **39**, 9153 (1989).

4. P. L. Gammel, L. F. Schneemeyer, J. V. Waszczak, and D. J. Bishop, Phys. Rev. Lett. **61**, 1666 (1988).

5. E. H. Brandt, Phys. Rev. Lett **63**, 1106 (1989).

6. E. H. Brandt, Int. J. Mod. Phys. **B 5**, 751 (1991).

7. E. H. Brandt, Rep. Prog. Phys. **58**, 1465 (1995).

8. M. A. Moore, Phys. Rev. B **39**, 136 (1989); A. Houghton, R. A. Pelcovits, and A. Sudbø, Phys. Rev. B **40**, 6763 (1989).

9. P. L. Gammel, L. F. Schneemeyer, and D. J. Bishop, Phys. Rev. Lett **66**, 953 (1991).

10. D. S. Fisher, M. P. A. Fisher, and D. A. Huse, Phys. Rev. B **43**, 130 (1991).

11. T. Giamarchi and P Le Doussal, Phys. Rev. Lett. **72,** 1530 (1994); Phys. Rev. B **52,** 1242 (1995) and references therein.

12. M. Gingras and D. A. Huse, Phys. Rev. B **53**, 15193 (1996).

13. T. Giamarchi and Pierre Le Doussal, *Spin Glasses and Random Fields*, A. P. Young ed., (World Scientific, Singapore, 1998).

14. T. Natterman and S. Scheidl, Adv. Phys. **49**, 607 (2000).

15. G. Blatter, V. B. Geshkenbein, A. Larkin, and H. Nordborg, Phys. Rev. B **54**, 72 (1996).

16. G. Blatter and B. Ivlev, Phys. Rev. B **50**, 10272 (1994); L. I. Glazman and A. E. Koshelev, Phys. Rev. B **43**, 2835 (1991).

17. R. B. van Dover, L. F. Schneemeyer, E. M. Gyorgy, and J. V. Waszczak, Phys. Rev. B **39**, 4800(R) (1989); D. E. Farrell, J. P. Rice, and D. M. Ginsberg, Phys. Rev. Lett. **67**, 1165 (1991); H. Safar, P. L. Gammel, D. A. Huse, D. J. Bishop, J. P. Rice, and D. M. Ginsberg, Phys. Rev. Lett. **69**, 824 (1992); W. Kwok, J. Fendrich, S. Fleshler, U. Welp, J. Downey, and G. W. Crabtree, Phys. Rev. Lett. **72**, 1092 (1994).

18. E. Zeldov, D. Majer, M. Konczykowski, V. B. Geshkenbein, V. M. Vinokur, and H. Shtrikman, Nature **375,** 373 (1995).

19. A. Schilling, R. A. Fisher, N. E. Phillips, U. Welp, D. Dasgupta, W. K. Kwok, and G. W. Crabtree, Nature **382,** 791 (1996).

20. S. Colson, M. Konczykowski, M. B. Gaifullin, Y. Matsuda, P. Gierlowski, M. Li, P. H. Kes, and C. J. van der Beek, Phys. Rev. Lett. **90**, 137002 (2003).

21. S. S. Banerjee, A. Soibel, Y. Myasoedov, M. Rappaport, E. Zeldov, M. Menghini, Y. Fasano, F. de la Cruz, C. J. van der Beek, M. Konczykowski, and T. Tamegai, Phys. Rev. Lett. **90**, 087004 (2003).

22. M. J. W. Dodgson, A. E. Koshelev, V. B. Geshkenbein, and G. Blatter, Phys. Rev. Lett. **84**, 2698 (2000).

23 A. Crisan, S. J. Bending, Z. Z. Li, and H. Raffy, Supercond. Sci. Technol. **24** 115001 (2011).

24. G. Shaw, P. Mandal, S. S. Banerjee, and T. Tamegai, New J. Phys. **14,** 083042 (2012).

25. J. P. Lv, and Q. H. Chen, Phys. Rev. B **78**, 144507 (2008).

26. Q. H. Chen, Q. M. Nie, J. P. Lv, and T. C. Au Yeung, New J. Phys. **11,** 035003 (2009).

27. N. Avraham et. al., Nature **411**, 451 (2001).

28. Y. Yin, M. Zech, T. L. Williams, X. F. Wang, G. Wu, X. H. Chen, and J. E. Hoffman Phys. Rev. Lett. **102**, 097002 (2009).

29. C. L. Song, Y. Yin, M. Zech, T. Williams, M. M. Yee, G. F. Chen, J. L. Luo, N. L. Wang, E. W. Hudson, and J. E. Hoffman, Phys. Rev. B **87**, 214519 (2013).

30. C. J. van der Beek et. al., Phys. Rev. B **81**, 174517 (2010).

31. L. Shan, Nat. Phys. **7**, 325 (2011).

32. H. K. Mak, P. Burger, L. Cevey, T. Wolf, C. Meingast, and R. Lortz, Phys. Rev. B **87**, 214523 (2013).

33. T. Taen, F. Ohtake, H. Akiyama, H. Inoue, Y. Sun, S. Pyon, T. Tamegai, and H. Kitamura, Phys. Rev. B **88**, 224514 (2013).

34. P. Mandal, D. Chowdhury, S. S. Banerjee, and T. Tamegai, Rev. Sci. Instrum. **83**, 123906 (2012).

35. A. A. Polyanskii, D. M. Feldmann, and D. C. Larbalestier, *Magneto-optical characterization techniques Handbook of Superconducting Materials,* 2 ed D. A. Cardwell and D. S. Ginley (Bristol: Institute of Physics Publishing, 2003).

36. A. Soibel, E. Zeldov, M. Rappaport, Y. Myasoedov, T. Tamegai, S. Ooi, M. Konczykowski, and V. B. Geshkenbein, Nature **406,** 282–7 (2000); A. Soibel, Y. Myasoedov, M. L. Rappaport, T. Tamegai, S. S. Banerjee and E. Zeldov, Phys. Rev. Lett. **87,** 167001 (2001).

37. M. Daeumling, J. M. Seuntjens, and D. C. Larbalestier, Nature (London) **346**, 332-335 (1990); M. F. Goffman, J. A. Herbsommer, F. de la Cruz, T. W. Li, and P. H. Kes, Phys. Rev. B **57**, 3663 (1998) and references therein.

38. T. Tamegai, Y. Iye, I. Oguro, and K. Kishio, Physica C **213**, 33 (1993); B. Khaykovich, E. Zeldov, D. Majer, T. W. Li, P. H. Kes and M. Konczykowski, Phys. Rev. Lett. **76**, 2555 (1996).

39. D. Ertas and D. R. Nelson, Physica C **272**, 79 (1996).

40. B. Khaykovich, M. Konczykowski, E. Zeldov, R. A. Doyle, D. Majer, P. H. Kes, and T. W. Li, Phys. Rev. B **56**, 517(R) (1997).

41. K. Deligiannis, P. A. J. de Groot, M. Oussena, S. Pinfold, R. Langan, R. Gangon, and L. Taillefer, Phys. Rev. Lett. **79**, 2121 (1997); H. Küpfer, Th. Wolf, C. Lessing, A. A. Zhukov, X. Lancon, R. Meier-Hirmer, W. Schauer, and H. Wühl, Phys. Rev. B **58**, 2886 (1998); S. Okayasu and H. Asaoka, Physica C **317-318**, 633 (1999).

42. D. Giller, A. Shaulov, R. Prozorov, Y. Abulafia, Y. Wolfus, L. Burlachkov, Y. Yeshurun, E. Zeldov, V. M. Vinokur, J. L. Peng, and R. L. Greene, Phys. Rev. Lett. **79**, 2542 (1997).

43. T. Nishizaki, T. Naito and N. Kobayashi, Phys. Rev. B **58**, 11169 (1998); Physica C **282-287**, 2117 1997; **317-318**, 645 (1999).

44. D. Giller, A. Shaulov, Y. Yeshurun, and J. Giapintzakis, Phys. Rev. B **60**, 106 (1999).

45. S. Kokkaliaris, P. A. J. de Groot, S. N. Gordeev, A. A. Zhukov, R. Gagnon, and L. Taillefer, Phys. Rev. Lett. **82**, 5116 (1999).

46. S. S. Banerjee *et al.*, Phys. Rev. B **62**, 11838 (2000).

47. S. Mohan, J. Sinha, S. S. Banerjee, and Y. Myasoedov, Phys. Rev. Lett. **98**, 027003 (2007).

48. H. Yang, H. Q. Luo, Z. S. Wang, and H. H. Wen, Appl. Phys. Lett. **93**, 142506 (2008).

49. S. Demirdiş, C. J. van der Beek, S. Mühlbauer, Y. Su, and T. Wolf, J. Phys. Cond. Matt. **28**, 425701 (2016).

50. R. Prozorov, N. Ni, M. A. Tanatar, V. G. Kogan, R. T. Gordon, C. Martin, E. C. Blomberg, P. Prommapan, J. Q. Yan, S. L. Bud'ko, and P. C. Canfield, Phys. Rev. B **78**, 224506 (2008).

51. R. Prozorov, M. A. Tanatar, E. C. Blomberg, P. Promma-pan, R. T. Gordon, N. Ni, S. L. Bud'ko, and P. C. Canfield, Physica C **469**, 667 (2009).

52. R. Prozorov, M. A. Tanatar, N. Ni, A. Kreyssig, S. Nandi, S. L. Bud'ko, A. I. Goldman, and P. C. Canfield, Phys. Rev. B **80**, 174517 (2009).

53. Y. Nakajima, Y. Tsuchiya, T. Taen, T. Tamegai, S. Okayasu, and M. Sasase, Phys. Rev. B **80**, 012510 (2009).

54. C. P. Bean, Rev. Mod. Phys. **36**, 31 (1964).

55. H. P. Wiesinger, F. M. Sauerzopf, and H. W. Weber, Physica C **203**, 121 (2003).

56. E. Zeldov, A. I. Larkin, V. B. Geshkenbein, M. Konczykowski, D. Majer, B. Khaykovich, V. M. Vinokur, and H. Shtrikman, Phys. Rev. Lett. **73**, 1428 (1994).

57. B. Khaykovich, M. Konczykowski, K. Teitelbaum, E. Zeldov, H. Shtrikman, and M. Rappaport, Phys. Rev. B **57**, 14088(R) (1998).

58. E. H. Brandt, Phys. Rev. B **59**, 2269 (1999).

59. R. J. Wijngaarden, K. Heeck, H. J. W. Spoelder, R. Surdeanu, and R. Griessen, Physica C **295**, 177 (1998).

60. E. H. Brandt, G. P. Mikitik, and E. Zeldov, J. Exp. Theor. Phys. **117,** 439 (2013).

61. P. E. Goa, H. Hauglin, A. A. F. Olsen, D. Shantsev, and T. H. Johansen, Appl. Phys. Lett. **82**, 79 (2003).

62. U. Welp, R. Xie, A. E. Koshelev, W. K. Kwok, H. Q. Luo, Z. S. Wang, G. Mu, and H. H. Wen, Phys. Rev. B **79**, 094505 (2009).

63. A. Gurevich, Rep. Prog. Phys. **74**, 124501 (2011).

64. U. Welp, S. Fleshier, W. K. Kwok, R. A. Klemm, V. M. Vinokur, J. Downey, B. Veal, and G. W. Crabtree, Phys. Rev. Lett. **67**, 3180 (1991).

65. Z. Tesanovic, L. Xing, L. Bulaevskii, Q. Li, and M. Suenaga, Phys. Rev. Lett **69**, 3563 (1992).

66. Q. Li, K. Shibutani, M. Suenaga, I. Shigaki, and R. Ogawa, Phys. Rev B **48**, 9877 (1993).

67. A. Wahl, V. Hardy, F. Warmont, A. Maignan, M. P. Delamare, and Ch. Simon, Phys. Rev. B **55**, 3929 (1997).

68. B. Rosenstein, B. Ya. Shapiro, R. Prozorov, A. Shaulov, and Y. Yeshurun, Phys. Rev. B **63**, 134501 (2001).

69. S. Salem-Sugui, L. Ghivelder, A. D. Alvarenga, J. L. Pimentel, H. Luo, Zh. Wang, and H. H. Wen, Phys Rev. B **80**, 014518 (2009).

70. M. Tinkham*, Introduction to Superconductivity*, second edition, (Mc Graw-Hill, Inc., New York, 1996).
71. N. D. Mermin and H. Wagner, Phys. Rev. Lett. **17,** 1133 (1966).

72. P. Sen *et al*, Phys. Rev. Lett. **80**, 4092 (2001); S. S. Banerjee *et. al*. Phys. Rev. Lett. **93**, 097002 (2004).

# Planar pinning induced, lowering of vortex dimensionality and low field melting in a single crystal of $Ba_{0.6}K_{0.4}Fe_2As_2$: Supplementary Material

Ankit Kumar[1], Sayantan Ghosh[1], Tsuyoshi Tamegai[2], S. S. Banerjee[1,‡]

[1]Department of Physics, Indian Institute of Technology, Kanpur-208016, India

[2]Department of Applied Physics, The University of Tokyo, Hongo, Bunkyo-ku, Tokyo 113-8656, Japan

## Magneto-optical image showing the directed nature of vortex penetration through defect planes

Due to the presence of extended defects in the sample, the $B_{c1}$ is lowered locally over these regions. To demonstrate this, we show below conventional magneto – optical image (viz., an image representing $B_z(x,y)$ distribution across the sample) associated with penetrating flux in a single crystal of $Ba_{0.6}K_{0.4}Fe_2As_2$ superconductor. This single crystal has been chosen from the same crystal batch as discussed in the paper. The sample has identical $T_c$ and pinning properties as that discussed in the paper. Furthermore, we have chosen the sample with shape and dimensions close to that of the sample reported in the paper. To image the penetrating vortex state, we perform conventional MO imaging on the chosen sample which is zero field cooled, ZFC (viz., cooled below $T_c$ in nominally zero field and then the field is applied). Note the image shown below is a conventional MO image of a sample, which has been ZFC at 28 K and then a field of 24 Oe is applied. We see clearly preferential penetration of flux along with the linear defects present in the sample (similar to the ones discussed in the context of results shown in Fig. 3, 5 of our paper). The flux penetrates the sample from these linear defects in the sample at fields which are much lower than the bulk $B_{c1}(T)$. Hence, locally, the $B_{c1}$ over the regions with defect planes is much lower than the bulk. Note the contrast in this image, has been artificially enhanced to show the bright penetration of flux preferentially over the defects.

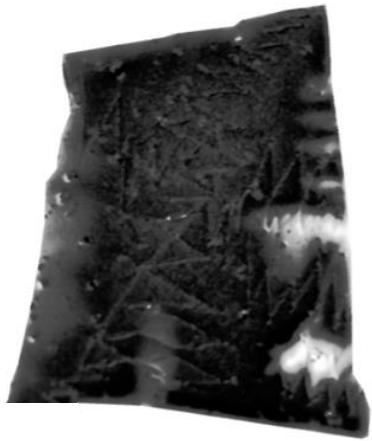

**FIG. 1**. A conventional MO image taken for a ZFC state at 28 K and at an applied field of 24 Oe ($<B_{c1}$). Note these images are not DMO images. We also reiterate they are taken for the ZFC state (flux is penetrating).

## Determination of anisotropy from fitting $H_{c2}(\theta)$ using GL equation

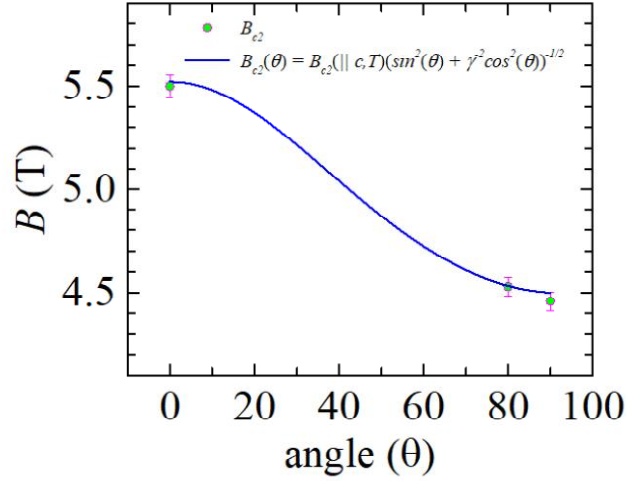

**FIG. 2**. In Fig. 2 we have plotted $B_{c2}(T)$ for different orientation of the sample w.r.t the applied field taken at 35 K. The $B_{c2}$ for the angles fits with the well-known GL expression $B_{c2}(\theta) = B_{c2}(||c,T)(sin^2\theta + \gamma^{-2}cos^2\theta)^{-1/2}$ where $\gamma$ is the sample anisotropy [1]. The fit gives us an estimate of the sample anisotropy to be ~1.22±0.11, which is close to the values reported in the literature.

### Determination of $\lambda_0$

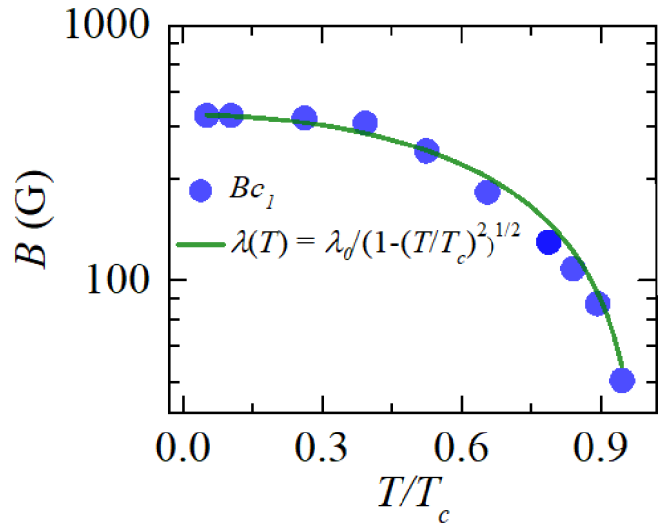

**FIG. 3**. Determination of $\lambda_0$ from the $B_{c1}$ plot.

In Fig. 3 we have shown the $B_{c1}$ data obtained from the magnetization $M(B)$ measurement using SQUID magnetometer. The above data has been taken when the sample has been zero field cooled (ZFC), and then the magnetic field was increased. We have fitted the obtained $B_{c1}(T)$ behaviour using $\lambda(T) = \lambda_0 /(1 - (T/T_c)^2)^{1/2}$. We have obtained the $\lambda_0 \approx 200$ nm. The $B_{c1}$ measured is the bulk lower critical field.

**Supplementary Material Video 1**. In supplementary material video 1, we have shown the propagation of vortex melting across the sample. The images are taken at 30.2 K temperature, and the field values have been mentioned in the video. Images used for making the video have been colored for better representation of propagation of vortex melting across the sample. In some of the images, e.g., 20, 24, 30 G images the zig-zag formation at the left edge of the sample is coming from the in-plane magnetization related to magnetic domain present on the magneto-optical indicator directly put on the sample for Faraday rotation. Blue color region in the video represents vortex density region $\delta B_z = \delta B_a = 1$ G (vortex solid; VS) and yellow region represents enhanced vortex density region with $\delta B_z > 1$ G (vortex liquid; VL).

Corresponding Author: ‡satyajit@iitk.ac.in

1. M. Tinkham, *Introduction to Superconductivity*, second edition, (Mc Graw-Hill, Inc., New York, 1996).